# Young's double-slit interference with single hard X-ray photons

TIMUR E. GUREYEV,[1,2,*] CHRIS J. HALL,[3] BENEDICTA ARHATARI,[3,4] DANIELE PELLICCIA,[5] ALALEH AMINZADEH,[6] KONSTANTIN M. PAVLOV,[7,8,2] AND HARRY M. QUINEY[1]

[1]*School of Physics, The University of Melbourne, Parkville, VIC 3010, Australia*
[2]*School of Physics and Astronomy, Monash University, Clayton, VIC 3800, Australia*
[3]*Australian Synchrotron, ANSTO, Clayton, VIC 3168, Australia*
[4]*School of Computing, Engineering and Mathematical Science, La Trobe University, Bundoora, VIC 3086, Australia*
[5]*Instruments & Data Tools Pty Ltd, Rowville, VIC 3178, Australia*
[6]*Research School of Physics, Australian National University, Canberra, ACT 2600, Australia*
[7]*School of Physical and Chemical Sciences, University of Canterbury, Christchurch 8041, New Zealand*
[8]*School of Science and Technology, University of New England, Armidale 2351, Australia*
*[*]timur.gureyev@unimelb.edu.au*

**Abstract:** Double-slit interference experiments using monochromatic hard X-rays with the energy of 25 keV are presented. The experiments were performed at a synchrotron source with a distance of 110 m between the interferometer and the detector to produce an interference pattern with a sufficiently broad period that could be adequately sampled by a photon-counting detector with 75 micrometre pixels. In the single-particle version of the experiment, over one million image frames with a single registered photon in each one were collected. The sum of these frames showed a clear presence of the interference pattern with the expected period. Subsequent analysis provided an objective estimation of the minimal number of detected photons required to determine, in accordance with the Rose criterion, the presence of the photon interference. Apart from a general theoretical interest, these investigations were aimed at exploring the possibility of medical X-ray phase-contrast imaging in photon-counting regime at minimal radiation doses.

## 1. Introduction

The classical Young's double-slit (Y2s) experiment was originally carried out by Thomas Young in 1801 using visible light [1]. In this and many similar subsequent experiments, the observed image pattern was interpreted as the result of interference of the two waves emerging from the individual slits of the interferometer, which led to the development of the notion of partial coherence in optics [2-4]. The single-particle version of the Y2s experiment (spY2s), where the interference pattern is produced by sequentially registering one quantum of radiation or matter waves at a time, was proposed in the first half of the 20th century. The spY2s experiment was successfully carried out with visible light photons [5], electrons [6], neutrons [7] and other forms of radiation and matter waves. Richard Feynman called the single-particle interference "a phenomenon … which has in it the heart of quantum mechanics. In reality, it contains the only mystery <of quantum mechanics>" [8]. In 2002, spY2s was voted "the most beautiful experiment" in physics by readers of Physics World [9].

While the classical Y2s experiment with hard X-rays was performed by Leitenberger *et al.* [10], it appears that the spY2s experiment has not been performed with X-rays yet. We describe the results of such an experiment with 25 keV X-rays in the present paper, utilising a photon-counting X-ray detector and an interferometer with two 3 μm wide slit-like openings in a 70 μm thick layer of gold. We also carry out a quantitative analysis of the progressively accumulated single-photon image frames with the goal of finding the minimal total number of detected photons which would allow one to reliably detect the presence of the double-slit interference pattern. In all the previously reported spY2s experiments, it was initially impossible to discern the interference pattern in the seemingly random distributions of a few detected quanta of radiation or matter, until the number of registered quanta would become large enough to form an interference pattern noticeable by eye. However, to the best of our knowledge, the question about an objective detection of the emergence of the interference pattern, with a minimal total number of detected quanta, has not been addressed yet. Apart from the general theoretical interest, such a question can be particularly pertinent in the context of future biomedical applications of very-low-dose X-ray imaging or, indeed, in any similar experimental setup, where the total number of photons, or other particles used for imaging, is small. Note that the need to detect the interference pattern and measure its characteristics, such as, for example, the period and the fringe visibility, with the minimal total number of photons, may arise either due to the requirement to minimise the radiation dose and the associated damage imparted to the sample, e.g. in the context of biomedical imaging, or due to the intrinsic scarcity of available photons, as in some astronomical applications.

As the basics of both the classical Y2s and spY2s experiments have been extensively described and debated in scientific and popular literature over the years [4,8,9], we omit the generic details here and proceed directly with a technical outline relevant to our experimental setup in Section 2. The results of our Y2s experiments are described in Section 3. The question of objective detection of the double-slit interference with a minimal number of photons is considered in Section 4. Finally, Section 5 contains some concluding remarks.

## 2. Design of the experiments

The one-dimensional (1D) photon fluence distribution in the detector plane, $I(y)$, produced by a Y2s interferometer, can be described by the following expression [4,5,10]:

$$I(y) = 2I_0 \left[ \frac{\sin(2\pi y / A)}{2\pi y / A} \right]^2 \left[ 1 + V \cos(2\pi y / a) \right] + I_b,$$ (1)

where $y$ is the coordinate orthogonal to the slit direction. The quantity $I_0$ denotes the uniform incident 1D fluence and $A = 2\lambda R / d$ is the "period" of the corresponding single-slit diffraction envelope. The parameter $a = \lambda R / D$ in eq.(1) is the period of the double-slit interference pattern, $\lambda$ is the wavelength (the illumination is assumed to be monochromatic here), $R$ is the slits-to-detector distance, $d$ is the width of each slit, $D$ is the separation distance between the slits, $V = \sin(\pi\xi) / (\pi\xi)$ is the fringe visibility, $\xi = sD / (\lambda L)$, $s$ is the size of the light source along the $y$ coordinate and $L$ is the source-to-slits distance. Finally, $I_b$ is the background fluence which may be produced by photons other than those that reached the detector directly after passing through the slits. The detected background may also contain other contributions, such as e.g. from cosmic rays or detector dark current. We will assume that each 1D fluence distribution like the one described by eq.(1), is the result of integration over $x$ of a two-dimensional (2D) photon fluence distribution, $I(x, y)$, registered by a 2D detector. In our experiment (described below) the slits were oriented along the horizontal direction, $x$, hence the choice of notation in eq.(1) with the variable $y$ running in the vertical direction orthogonal to the slits.

In the considered setup of Y2s interferometry, the fringe visibility coincides with the degree of spatial coherence of the two interfering waves originated from the individual slits [2-4]. When $V = 0$, eq.(1) describes the pattern produced when the complex amplitudes, corresponding to the waves transmitted through each slit, cannot interfere. In other words, this pattern is just a sum of intensities produced by the two slits separately. As a transverse shift of a structure in the object plane only leads to an addition of a linear phase term in the complex amplitude in a diffraction plane located in the Fraunhofer region, such shifts do not change the intensity distribution of the diffraction pattern. Therefore, each of the two transversally shifted identical slits produce the same diffraction pattern. Consequently, the sum of the two diffraction patterns produced by two incoherent slits when $V = 0$, is the same as the pattern produced by a single slit with twice the number of photons passed through it. Note that the parameter $\xi$ in the visibility expression, $V(\xi) = \sin(\pi\xi)/(\pi\xi)$, can be expressed as $\xi = sR/(aL) = s'/a$, where $s' \equiv sR/L = s(M-1)$ is the magnified source size in the image plane, with $M = (R+L)/L$ being the conventional geometric magnification coefficient of the imaging setup. Therefore, the visibility $V$ of the double-slit diffraction pattern is determined by the ratio of the magnified source size to the period of the double-slit interference pattern. If $s' >> a$, the visibility is close to zero. In this case, the magnified source effectively smears the double-slit diffraction pattern, making it indistinguishable from a single-slit pattern. On the other hand, when $s' << a$, the visibility is high, i.e. close to one. These considerations are central for our experimental design described below.

In order to obtain a high-quality double-slit diffraction pattern, the following two conditions should be satisfied in a Y2s experiment. These conditions guarantee that the double-slit diffraction pattern can be easily distinguished from the corresponding single-slit pattern and that it can also be adequately sampled by the detector with a given spatial resolution.

*Condition 1.* The period of the double-slit diffraction pattern should be broad enough to allow for an adequate sampling by a detector with a given spatial resolution $\Delta$. Shannon sampling (two points per period) imply $a \geq 2\Delta$, and the two-times over-sampling of noisy data may require at least $a \geq 4\Delta$.

*Condition 2.* Visibility of the two-slit diffraction, $V(\xi) = \sin(\pi\xi)/(\pi\xi)$, should be high enough for the diffraction pattern to be detectable. This can be ensured, for example, by requiring that $\xi < 0.6$, which would result in the visibility higher than $V(0.6) \cong 0.5$.

Satisfying these conditions in a spY2s experiment with hard X-rays presents a number of technical challenges that are specific to the high penetrating power of hard X-rays and their short wavelengths. For a spY2s experiment, a photon-counting X-ray detector is needed. The spatial resolution of such detectors often coincides with the pixel size, which is typically between 50 μm and 100 μm for most modern photon-counting X-ray detectors [11]. The present experiment was designed for the Eiger2-3MW detector [12] which has the pixel size $\Delta = 75$ μm. Condition 1 with two-times oversampling then implies that the period, $a$, of the double-slit diffraction pattern should be at least 300 μm.

Our spY2s experiment was designed for the Imaging and Medical beamline (IMBL) of the Australian Synchrotron [13]. The practical range of monochromatic X-ray energies at IMBL is approximately between 20 keV and 120 keV. According to eq.(1), the period, $a = \lambda R/D$, of the double-slit diffraction pattern is proportional to the radiation wavelength. Therefore, in order to have a larger period satisfying Condition 1, it was preferable to use longer wavelengths, i.e. lower X-ray energies. We chose to use monochromatic X-rays with the energy $E = 25$ keV ($\lambda \cong 0.5$ Å) selected by a double bent-crystal monochromator working in the Laue geometry [13]. It was also necessary to ensure that the interferometer-to-detector distance $R$ would be sufficiently large and the slit separation distance $D$ would be sufficiently small. As the slit separation is defined as the distance between the central lines of the slits, the slit width $d$ is always smaller or equal than $D$. This presents a particular challenge in the case of hard X-rays, since a sufficiently thick layer of highly absorbing material is required outside the (narrow) slits in order to stop the energetic X-rays. For example, a 60 μm thick layer of gold has slightly less than 1% transmissivity for 25 keV X-rays. Therefore, for example, for 3 μm slits, the anisotropy of the required profile (the ratio of the slit width to the height of the slit walls) is $3:60 = 1:20$. This magnitude of anisotropy appears to be close to the limits of the current micro-manufacturing technology [14]. We chose the width $d = 3$ μm for the slits of a Y2s interferometer device designed for this experiment, with three slit pairs with the distances $D$ between the slits equal to 6, 12 and 18 μm in the different pairs.

In order to maximize the interferometer-to-detector distance $R$, we decided to place the interferometer in IMBL's hutch 2B and the detector in hutch 3B [13], resulting in the source-to-interferometer distance $L = 36$ m and the interferometer-to-detector distance $R = 110$ m. With this design, the width of the central peak of the single-slit diffraction pattern was $A = 2\lambda R / d \cong 3667\,\mu\text{m}$ and the double-slit diffraction periods $a = \lambda R / D \cong 5500\,\mu\text{m}^2 / D$ were equal to approximately 917 μm, 458 μm and 306 μm for the $D$ values of 6, 12 and 18 μm, respectively. Even the smallest of these values, $a = 306$ μm, was larger than than four detector pixels, $4 \times 75\,\mu\text{m} = 300\,\mu\text{m}$, hence Condition 1 was going to be satisfied.

Regarding Condition 2, the source size at IMBL is 40 μm (FWHM) in the vertical and 800 μm (FWHM) in the horizontal direction [13]. For this experiment, the interferometer was planned to be positioned with the slits running in the horizontal direction, so that the smaller (vertical) source size could be effectively utilised. With the interferometer in hutch 2B and the detector in hutch 3B, the geometrical magnification factor was $M = (R + L) / L \cong 4.06$ and the visibility was $V(\xi) = \sin(\pi\xi) / (\pi\xi)$ with $\xi = s(M - 1) / a \cong 0.022D$, with $D$ expressed in microns. This resulted in the visibility values of 0.97, 0.89 and 0.76 for the $D$ values of 6, 12 and 18 μm, respectively, satisfying Condition 2.

It follows from the above considerations that a Y2s experiment could be realistically conducted at IMBL with the interferometer in hutch 2B and the detector in hutch 3B, using a detector with the spatial resolution of 75 μm. Figure 1 shows simulated double-slit diffraction patterns with the chosen design parameters.

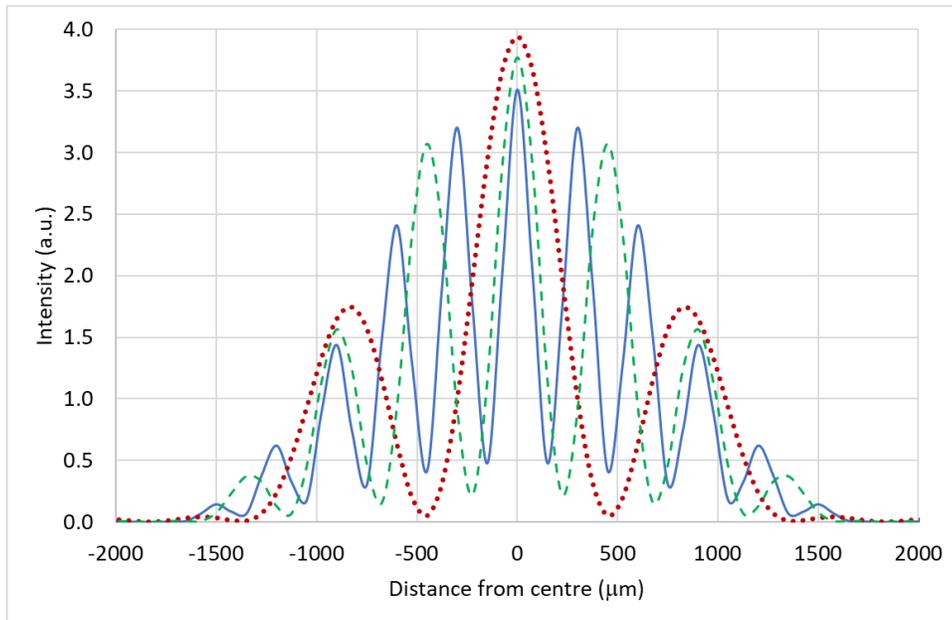

**Fig. 1.** Theoretical plot of the double-slit diffraction pattern, $I(y)$, as defined by eq.(1) with $\lambda \cong 0.5$ Å ($E \cong 25$ keV), $L = 36$ m, $R = 110$ m, $s = 40$ μm, $d = 3$ μm, $I_0 = 1$, $I_b = 0$. Solid blue line corresponds to the slit separation distance $D = 18$ μm, dashed green line - to $D = 12$ μm, and the dotted red line - to $D = 6$ μm.

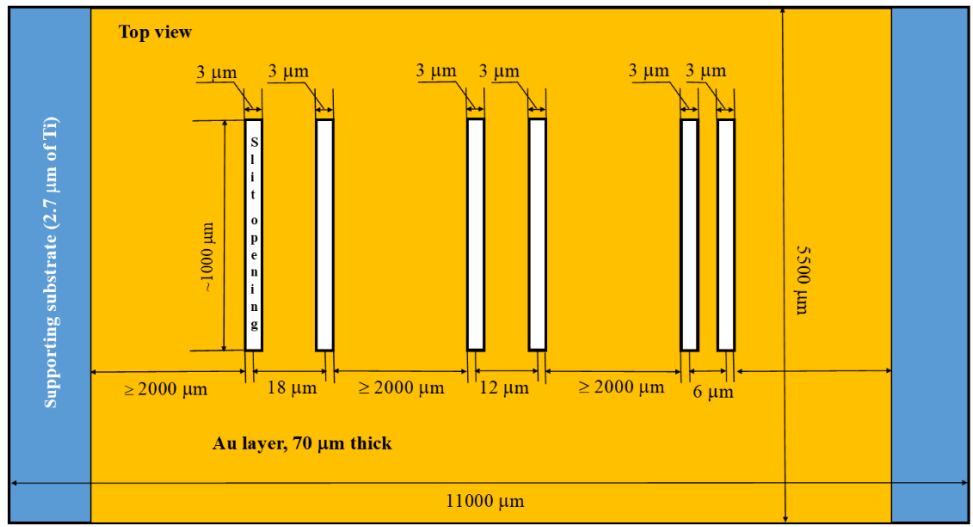

**Fig. 2.** The design of a Y2s interferometer device with three pairs of slits (openings), with a different slit separation in each pair, made in a plate covered with a layer of gold.

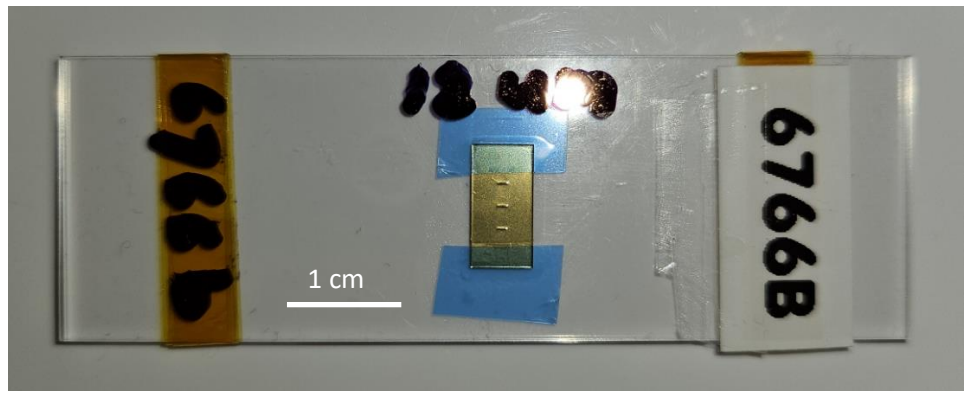

**Fig. 3.** Photograph of the interferometer device mounted between two microscope glass slides in preparation for the experiment. The three pairs of slits in the gold plating are visible. The slit pair with the widest (18 μm) slit separation is at the top position.

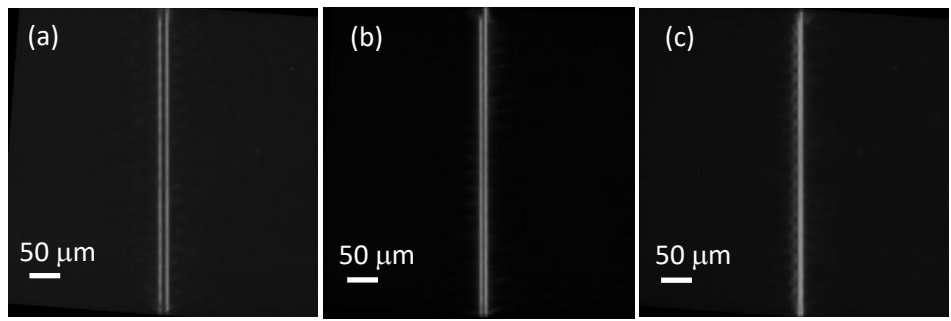

**Fig. 4.** "Contact" X-ray images of the three slit pairs of the Y2s interferometer device shown in Fig. 3: (a) $D$ = 18 μm, (b) $D$ = 12 μm, (c) $D$ = 6 μm.

The practical design of the double-slit interferometer device used in our experiment is shown in Fig.2. A description of technological processes used for manufacturing devices like this one can be found at the website of the company, Microworks GmbH (Karlsruhe, Germany) [14], which produced the device according to our specifications. Here we just note that the device was manufactured using a combination of X-ray lithography at the KARA Synchrotron [15] and gold electroplating. A photograph of the device mounted between two glass plates for the experiment is shown in Fig. 3. Figure 4 shows "contact" X-ray images of the three slit pairs collected at the MCT beamline of the Australian Synchrotron [16] with plane monochromatic 25 keV X-rays and a short distance (~16 cm) between the interferometer and the area detector. The detector pixel size here was ~$0.33 \times 0.33 \ \mu m^2$ and the effective spatial resolution was approximately 2 μm, which was affected by the X-ray source size [16].

### 3. Double-slit diffraction experiments

#### 3.1 Classical imaging regime

As planned at the design stage, the diffraction experiment was carried out at IMBL at the Australian Synchrotron [13] using plane monochromatic X-ray illumination with the energy $E = 25$ keV ($\lambda \cong 0.5$ Å) and monochromaticity $\Delta E / E \cong 0.001$. The source-to-interferometer distance was $L = 36$ m and the interferometer-to-detector distance was $R = 110$ m. In order to reduce the X-ray scattering in air along the beam path from the source to the interferometer and from the interferometer to the detector, most of the path was contained in evacuated beam-transfer pipes [13]. A photon-counting detector with the pixel size $\Delta = 75$ μm and a single-pixel point spread function was used.

Experimental diffraction profiles (2D diffraction patterns integrated over the $x$ coordinate) obtained for the $D = 18$ μm, $D = 12$ μm and $D = 6$ μm slit pairs are presented in Figs. 5, 6 and 7, respectively, alongside the corresponding theoretical profiles (with $s = 40$ μm). These experimental diffraction patterns were collected in the "classical" manner, i.e. at a relatively high photon flux. As one can see, the periodicity of the experimental diffraction profiles agrees well with the theoretical predictions, with some minor variations. In particular, we found that the envelope of the experimental diffraction profiles corresponded to the effective slit width of approximately 2 μm, instead of 3 μm. This was likely the result of a slight misalignment (pitch) of the interferometer with respect to the X-ray beam direction ($z$ axis). Note that with the slit walls height of 70 μm, the reduction of the effective slit width from 3 μm to 2 μm corresponds to a ~0.8 degree angle between the slit walls and the $z$ axis. This type of residual misalignment was possibly present in the experiment despite our best efforts to align the interferometer along the beam.

The visibility of the experimental diffraction curves was slightly lower compared to the theoretically predicted values. This was likely caused by the interferometer misalignment mentioned above, the effects of optical elements between the source and the interferometer, and also by the presence of the non-uniform background intensity, $I_b(y)$. The background was likely created by the X-ray transmission through thinner areas of the gold layer (especially, between the slits), as well as by the third harmonic of the monochromator with $E = 75$ keV X-rays (which penetrated the gold layer much stronger and also diffracted differently from the $E = 25$ keV X-rays), by the X-ray scattering from various components of the beamline, by cosmic rays and other factors. Nevertheless, the experimental data confirms that our interferometer device was functioning generally as expected in this "classical" regime.

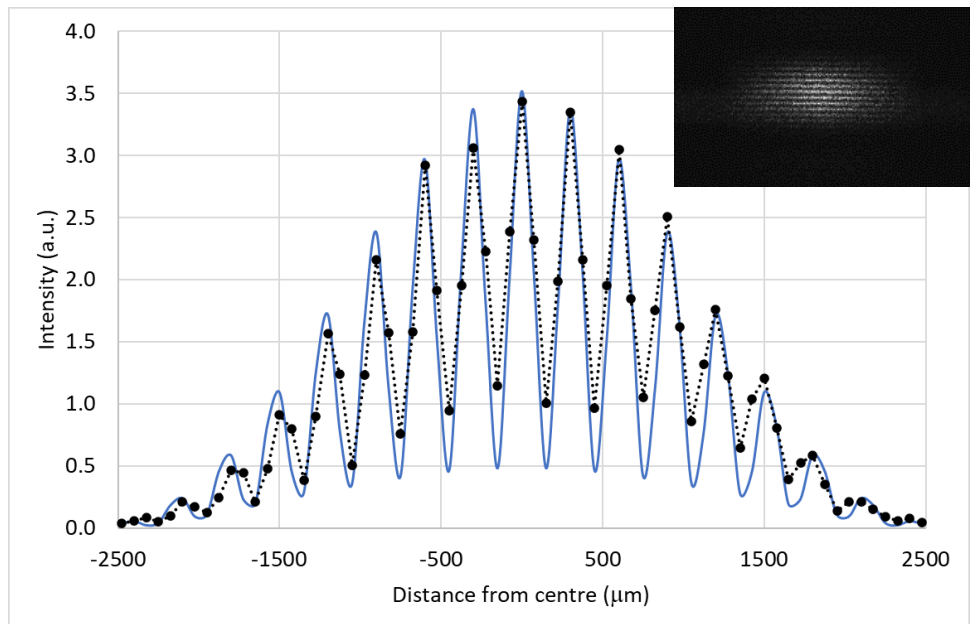

**Fig. 5.** Experimental diffraction profile (2D diffraction pattern integrated over the *x* coordinate) obtained for the $D = 18$ μm slit pair (dotted black line) and the corresponding theoretical diffraction profile with $d = 2$ μm and $D = 18$ μm (solid blue line). The insert shows the corresponding raw 2D experimental diffraction pattern.

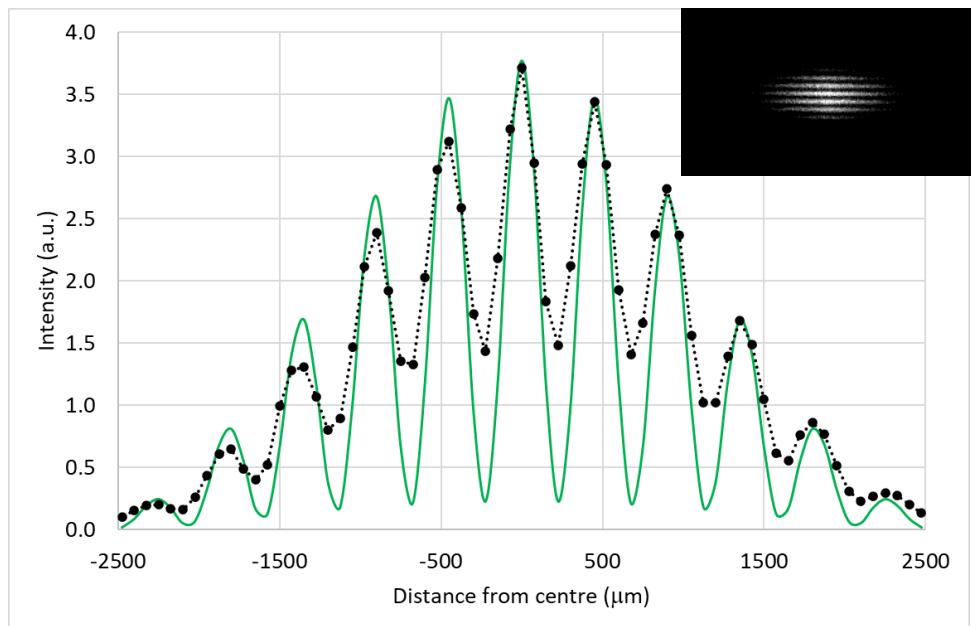

**Fig. 6.** Experimental diffraction profile (2D diffraction pattern integrated over the *x* coordinate) obtained for the $D = 12$ μm slit pair (dotted black line) and the corresponding theoretical diffraction profile with $d = 1.9$ μm and $D = 12$ μm (solid green line). The insert shows the corresponding raw 2D experimental diffraction pattern.

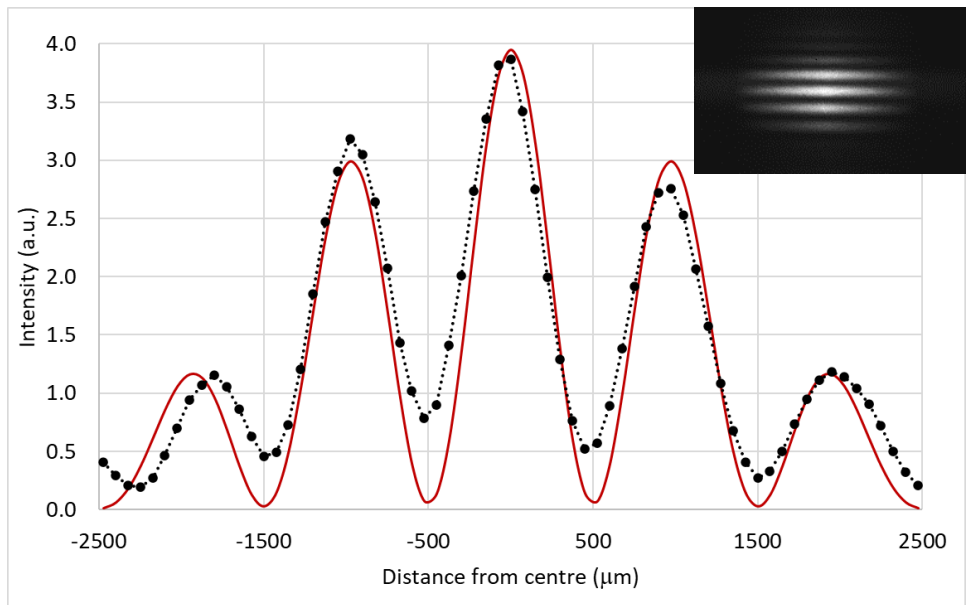

**Fig. 7.** Experimental diffraction profile (2D diffraction pattern integrated over the *x* coordinate) obtained for the $D = 6\,\mu\text{m}$ slit pair (dotted black line) and the corresponding theoretical diffraction pattern with $d = 1.6\,\mu\text{m}$ and $D = 5.5\,\mu\text{m}$ (solid red line). The insert shows the corresponding raw 2D experimental diffraction pattern.

### 3.2 Single-particle imaging regime

Only the slit pair with $D = 6\,\mu\text{m}$ was imaged in the single-particle regime, i.e. at a very low photon flux. Over $4.7{\times}10^6$ image frames, with $290 \times 277$ pixels each, were collected by running the Eiger2-3MW detector continuously for approximately 6.5 hours at 200 frames per second at a constant, very low X-ray flux. As the readout time on this detector was negligible due to its internal buffering, the exposure time for the individual image frames was 5 ms. The lowering of the flux was achieved by detuning the monochromator from its nominal Bragg peak position. In order not to overwhelm the file storage system on the computer where the images were saved, the frames were collected into 72 separate HDF5 files, each containing 65536 frames. We have checked the storage ring logs and the 72 individual HDF5 files that were collected during the low-flux image acquisition. One of these HDF5 files turned out to be defective and was discarded. Our analysis of the remaining 71 HDF5 files showed no abnormal fluctuations in the beam intensity during the whole acquisition period. In order to check the intensity fluctuations and the drift of the beam during the exposure, we created 71 images, each containing pixel-wise sums of the 65,536 image frames from the individual HDF5 files. The maximum intensity in any one pixel of these 71 integrated images was 13 photons, the minimum was zero. The total number of photons varied between 12371 and 17191 in the integrated images, with the standard deviation of approximately 967 photons. The standard deviation of the "centre of mass" of the integrated images was less than 0.5 of a pixel in both the horizontal and the vertical direction. After analysing all the collected frames, we sub-selected $1.04{\times}10^6$ frames, each containing exactly one registered photon event. Note that some of these registered "photon events" were possibly caused by cosmic rays and some corresponded to the third harmonic, i.e. to 75 keV X-ray photons. However, the majority of the registered photon events corresponded to single 25 keV X-ray photons, as demonstrated by the resultant diffraction patterns below.

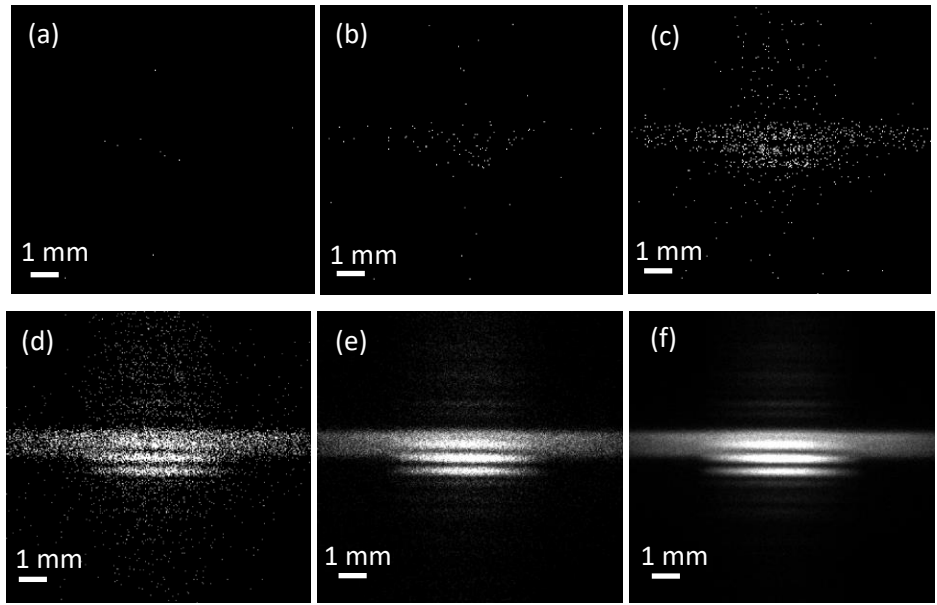

**Fig. 8.** Progressive accumulation of single-photon events. The images contain: (a) 10 photons; (b) $10^2$ photons; (c) $10^3$ photons; (d) $10^4$ photons; (e) $1.07 \times 10^5$ photons; (f) $1.04 \times 10^6$ photons. Incoherent background is visible primarily in the form of a horizontal stripe running across the images in the middle.

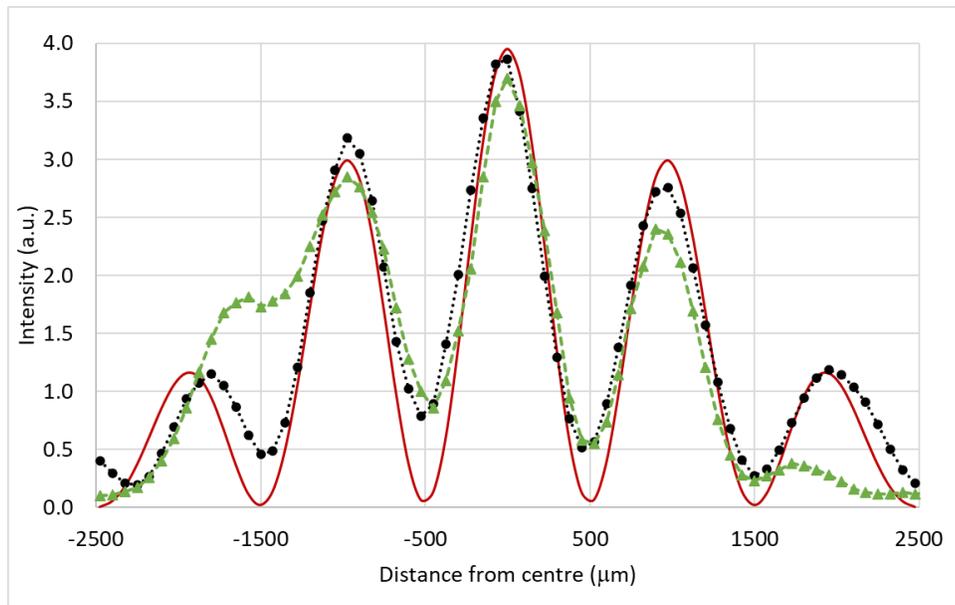

**Fig. 9.** Same as Fig. 7, but with the added accumulated single-photon interference profile with $1.04 \times 10^6$ photons, which is indicated by the dashed green curve with triangles. A contribution from the incoherent background to the latter curve is particularly evident in the interval between approximately $y = -2000$ μm and $y = -1000$ μm.

We progressively summed the $1.04 \times 10^6$ single-photon frames, with the results presented in Fig. 8. One can clearly see that the double-slit interference pattern became evident in these accumulated single-photon images. Some "incoherent" background was also visible, most prominently in the form of a horizontal band across the middle of the images. The fact that we did not see as much of a background in the high-flux images presented above, was due to the possibility to perform real-time fine-tuning of the alignment of optical elements of the beamline, when the diffraction patterns were clearly visible in real time.

## 4. Minimal number of photons required to reliably detect the presence of interference

Consider the problem of distinguishing a coherent diffraction pattern produced by two spatially separated slits (e.g. by a Y2s interferometer) from that produced by a single slit, with both patterns having the same total number of registered photons. The squared SNR of the difference between the two patterns, which is similar to the squared SNR of the "ideal observer" [17], can be expressed as follows [18]:

$$\text{SNR}^2[I, I_1] = 2 \int_{-\infty}^{\infty} \frac{[I(y) - I_1(y)]^2}{I(y) + I_1(y)} \, dy \,, \tag{2}$$

where $I(y)$ is defined by eq.(1) and $I_1(y)$ is defined by eq.(1) with $V = 0$. According to the arguments presented after eq.(1) above, eq.(2) also describes the ideal observer SNR for distinguishing between the diffraction patterns produced by two coherent slits, with the interference between the complex amplitudes corresponding to each slit, and two incoherent slits, with no interference.

Let us assume for simplicity that the background fluences are either negligible or can be subtracted from both the double-slit and the single-slit patterns before comparing them. Substituting the expressions for $I(y)$ and $I_1(y)$ given eq.(1) into eq.(2), and making the change of variables $y' = 2y/A$, we obtain:

$$\text{SNR}^2[I, I_1] = 2I_0 A \int_{-\infty}^{\infty} \left[ \frac{\sin(\pi y')}{\pi y'} \right]^2 \frac{V^2 \cos^2(\pi y' A/a)}{V \cos(\pi y' A/a) + 2} \, dy' \,. \tag{3}$$

We assume further that the two patterns are to be distinguished on the basis of their comparison within the central lobe of the single-slit diffraction pattern, $-1 \leq y' < 1$, and replace $\text{sinc}^2(\pi y')$ with 1 inside the latter interval. Accordingly, we replace eq.(3) with

$$\text{SNR}^2[I, I_1] \cong 2I_0 A V^2 \int_{-1}^{1} \frac{\cos^2(2\pi Dy'/d)}{V \cos(2\pi Dy'/d) + 2} \, dy' \,, \tag{4}$$

where we used the identity $A/a = 2D/d$. In order to simplify eq.(4) further, we also neglect the first term in the denominator, effectively assuming that $|V \cos(2\pi Dy'/d)| << 2$, i.e. assuming that the visibility is low. The resulting integral can be evaluated analytically with the result

$$\text{SNR}^2[I, I_1] \cong I_0 A V^2 \{1 + [d/(4\pi D)] \sin(4\pi D/d)\} \,. \tag{5}$$

As $d < D$ by construction, the second term in square brackets in eq.(5) is substantially smaller than unity. Omitting this smaller term, we obtain $\text{SNR}^2 \cong I_0 A V^2$, which can be re-written as

$$Q_S^2 \equiv \frac{\text{SNR}^2}{I_0 A} = \frac{\text{SNR}^2}{N} \cong V^2 \leq 1 \,, \tag{6}$$

where we normalized the incident intensity as $I_0 = N/A$, with $N$ being the total number of photons registered in the central lobe of the single-slit diffraction pattern. Equation (6) has a form of the noise-resolution uncertainty (NRU) [19]. It shows, in particular, that the ratio of $SNR^2$ to the total number of registered photons cannot exceed unity.

In order to establish a more direct correspondence between eq.(6) and the NRU presented in [19], we note first that the $SNR^2$ in the numerator of eq.(6) has an integral form, as it was integrated over the central lobe of the diffraction pattern. In contrast, the squared SNR utilised in [19] and related publications was "local", i.e. it corresponded to the average value of $SNR^2$ in a single detector pixel, $< SNR_{pix}^2 >$. In the setting considered above, we have $< SNR_{pix}^2 > \cong SNR^2 / M$, where $M$ is the number of pixels in the interval over which $SNR^2$ was calculated. Substituting this into eq.(6) we obtain

$$Q_S^2 = \frac{< SNR_{pix}^2 >}{I_0 \, \Delta}, \qquad (7)$$

where $\Delta = A / M$ corresponds to the pixel size. Equation (7) now coincides with the definition of the intrinsic imaging quality introduced in [19]. Note that the approximate expression for $Q_S$ given by eq.(6) equates it with the visibility, implying, in particular, that $Q_S \leq 1$. However, as shown in [19,20], the exact upper bound for $Q_S$ is slightly higher than unity, in general.

The quantity $Q_s$ was previously termed "intrinsic imaging quality" [19]; it describes the efficiency of the imaging setup in terms of utilisation of detected photons for extracting information about the imaged object. We see from eq.(6) that, in the case of Y2s interferometry, $Q_s$ is approximately equal to the fringe visibility. This is an intuitively logical result, since the extraction of information about the geometric parameters of an interferometer from the registered image directly depends on the fringe visibility. In particular, if the visibility is zero, no useful information about the interferometer can be extracted. The meaning of the word "useful" in this context is essentially determined by the definition of SNR in eq.(2), where the "useful signal" is proportional to the difference between the double slit and the single slit patterns. When the visibility is zero, the latter difference is also equal to zero, as can be seen from eq.(3).

One can use eq.(6) to predict the lowest number of photons required to reliably determine if the observed diffraction pattern corresponds to a coherent or an incoherent double-slit diffraction pattern. For this purpose, one can use, for example, the Rose criterion for distinguishability of patterns, $SNR \geq 5$ [21]. It follows from eq.(6) that $SNR_{Rose}^2 = 25 = V^2 N_{Rose}$. Note that the ensuing minimal number of photons, $N_{Rose} = 25 / V^2$, depends only on the fringe visibility (i.e. the degree of coherence), with no other parameters of the diffraction pattern affecting the result.

In our spY2s experiment, we had $s = 40$ μm, $D = 6$ μm, $\lambda = 0.5$ Å and $L = 36$ m, implying that $\xi = sD / (\lambda L) \cong 0.13$ and, hence, $V^2 = [\sin(\pi\xi) / (\pi\xi)]^2 \cong 0.94$. Therefore, in order to satisfy the Rose criterion for distinguishing between the double-slit and the single-slit diffraction patterns in our experiment, using the ideal observer SNR approach, it should be sufficient to accumulate just $N_{Rose} = 25 / 0.94 \cong 27$ photons in a diffraction pattern within the interval $-A / 2 \leq y < A / 2$, $A = 3667$ μm. Note, however, that this estimation is actually rather optimistic, due to the various approximations made in the derivation of eq.(6). It does not take into account some potentially detrimental factors, for example, the effect of the background intensities, which can increase the minimal number of photons required for distinguishing between the double-slit and the single-slit diffraction profiles.

In the spY2s experiment described in Section 3 above, we progressively collected over a million single-photon events in the diffraction pattern produced by the double-slit interferometer with $D = 6$ μm. We now want to estimate the minimal number of photons in the accumulated experimental diffraction pattern that would allow one to reliably detect the presence of the double-slit interference in accordance with eq.(2) and the Rose criterion. For this purpose, instead of measuring the difference between the coherent and the incoherent double-slit diffraction patterns, we measured the differences between the experimentally acquired diffraction patterns and the corresponding theoretical single-slit (incoherent) and double-slit (coherent) diffraction patterns. This allowed us to quantitatively evaluate the degree of similarity between the experimental diffraction patterns, containing a certain total number of photons, with the two theoretical patterns corresponding to the absence and the presence of the interference, respectively.

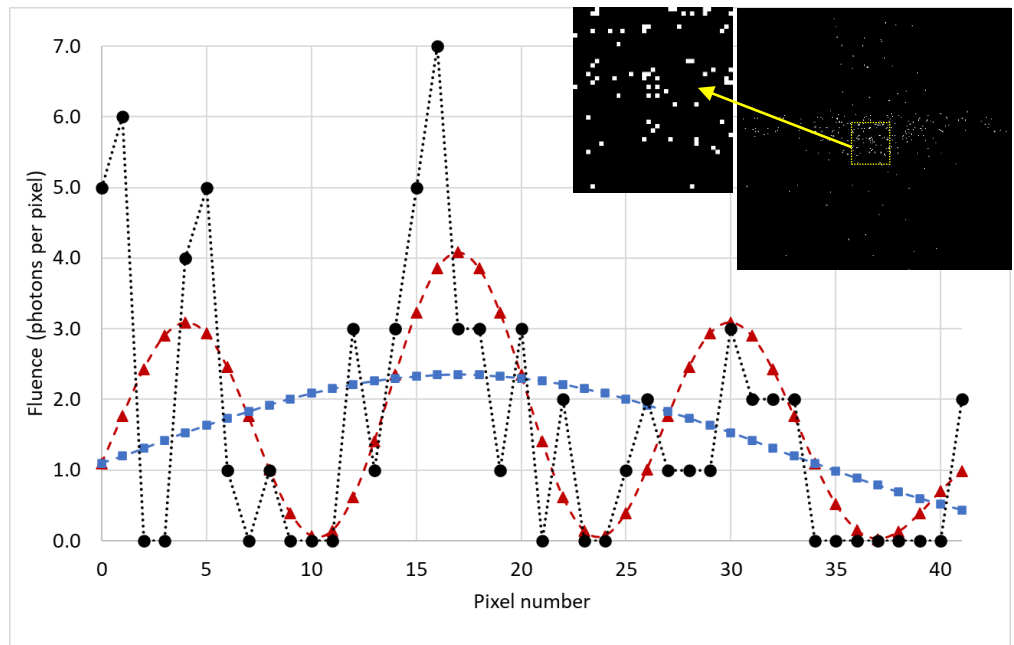

**Fig. 10.** Experimental diffraction profile (2D diffraction pattern integrated over the $x$ coordinate) obtained for the $D = 6\ \mu m$ slit pair, with $N = 70$ photons (dotted black line with circles), alongside the theoretical double-slit interference profile with $d = 1.6\ \mu m$ and $D = 5.5\ \mu m$ (dashed red line with triangles) and the corresponding single-slit profile (dashed blue line with squares), scaled in photons per pixel. The insert shows the raw 2D experimental diffraction pattern, with the yellow dotted rectangle outlining the sub-region where the measurements were performed.

Figure 10 shows the theoretical double-slit diffraction profiles with and without the interference for the parameters used in our spY2s experiment, together with the experimental diffraction profile obtained by summing the first 300 single-photon image frames. The final measurements were performed in the sub-region of $37 \times 42$ pixels of the summed image, which is indicated by the yellow dotted rectangle in Fig.10. This sub-region contained 70 photons. The $y$-dimension of the sub-region was slightly smaller than the value $A = 3667\ \mu m$ used in the above theoretical calculations, because the sub-region was truncated on the left-hand side in order to exclude the area with a particularly strong background (see Fig.9). We then calculated the experimental 1D diffraction profile (shown by the dotted black line in Fig.10) by integrating the registered intensity in the selected sub-region along the $x$ coordinate. We also scaled the corresponding theoretical double-slit and single-slit profiles to correspond to the same total number of photons, $N = 70$, over the measurement region. The $SNR^2$, numerically calculated according to eq.(2) for the difference between the resultant 1D experimental profile and the theoretical double-slit and single-slit profiles were, respectively, $\text{SNR}^2[I_{N=70}^{\text{experiment}}, I_{N=70}^{\text{theory}}] \cong 53.2$ and $\text{SNR}^2[I_{N=70}^{\text{experiment}}, I_{1,N=70}^{\text{theory}}] \cong 79.0$. Therefore, the difference between the squared SNR of the fits of the double-slit and the single-slit theoretical profiles by the experimental profile with 70 photons was approximately $79.0 - 53.2 = 25.8$. Hence, we can conclude that, under the conditions of our spY2s experiment, it was sufficient to collect approximately 70 photons in the diffraction pattern in order to reliably detect the presence of the double-slit interference. This result appears to be in a qualitative agreement with the visual appearance of Fig.10, in the sense that the presence of double-slit interference can just be discerned by eye in the experimental 1D diffraction profile shown in this figure.

Note that the estimated minimal number of detected photons, $N_{Rose} = 70$, required to reliably detect the presence of interference in our spY2s experiment, was significantly larger than the theoretical estimation, $N_{Rose} = 27$, obtained above for the ideal case. As mentioned, the latter estimation was overly optimistic, due to the various approximations used in the derivation of eq.(6). When we numerically calculated $\text{SNR}^2$ in accordance with the original eq.(2), for the theoretical profiles given by eq.(1) and the parameters of our spY2s experiment, the theoretical estimation for the minimal required number of photons became $N_{Rose} \cong 45$. The latter number is substantially closer to our experimentally obtained value of $N_{Rose} \cong 70$ than the initial rough estimation, $N_{Rose} = 27$. The remaining difference can be explained by the residual errors in the experimental alignment and the presence of the background intensity. We have eliminated some of the background by restricting the measurements to a sub-region where the background was less prominent, but some background intensity was certainly still present in the selected sub-region.

## 5. Conclusions

We have presented the results of a Y2s interference experiment in the classical (high flux) and the single-particle (one registered photon per detected image) versions, using monochromatic plane incident hard X-ray illumination with $E = 25$ keV. A unique feature of the experiment was the very long distance ($R = 110$ m) between the interferometer and the photon-counting detector. The long distance was required because of the short X-ray wavelength ($\lambda \cong 0.5$ Å), the relatively large width of the slits ($d = 3$ μm) and the size of the pixels ($\Delta = 75$ μm) of the available photon-counting detector. Both the classical and the single-particle interference patterns have been successfully acquired, with the parameters of the obtained patterns being in agreement with the known theory of the Y2s interferometry. As an additional result, we have also estimated theoretically and then measured experimentally the minimal number of photons that needs to be registered in a spY2s experiment in order to reliably detect the presence of the double-slit interference. Our approach was based on the theory of the "ideal observer" [17] and the Rose criterion for detection of image features. While the theoretically-estimated minimal number of photons under the conditions of the experiment was $N_{Rose} = 45$, our experimentally measured value was $N_{Rose} = 70$, with the difference attributable to the residual inaccuracies in the experimental setup and the presence of an incoherent background in the experimental diffraction patterns. The background was caused by a variety of factors, including the imperfection of the interferometer (thinner absorptive gold layer in some areas), the presence of the third X-ray harmonic (with $E = 75$ keV), the X-ray scattering from various components of the beamline and the contribution from cosmic rays.

This experiment was also aimed at testing the options for X-ray propagation-based phase-contrast imaging (PBI) with very long distances between the imaged sample and the detector [13,19]. In biomedical X-ray imaging, the minimization of radiation dose can be critically important. In this respect, our results about the minimal number of photons required to reliably distinguish certain features of an imaged object, provide a useful point of reference for future studies of quantum forms of medical X-ray imaging. The issue of medical image quality is also intrinsically linked to the problem of maximization of image contrast and SNR at a fixed dose. In PBI, the image contrast is linearly proportional to the sample-to-detector distance [19], hence, increasing the latter distance can be highly beneficial, in principle. However, the related detrimental factor is the image blurring due to the increase of the geometrically magnified source size with the sample-to-detector distance, which strongly reduces the contrast. In the present experiment we have utilised the smaller (vertical) dimension ($s \cong 40\ \mu m$) of the X-ray source at IMBL, which allowed us to achieve image fringe visibility (i.e. the contrast) close to unity. However, the horizontal source size at IMBL is much larger (800 μm). Consequently, when the interferometer was oriented with the slits running vertically and the diffraction profile variation was in the horizontal direction, no interference contrast could be observed in the considered experimental setup, as it was completely washed out by the large horizontal source size. In the future, it will be important to find ways to mitigate the effect of the geometrically magnified X-ray source on the image contrast, also known as the penumbral effect, in both the horizontal and the vertical directions, in order to be able to acquire useful 2D or 3D (in a tomographic setup) PBI images at long sample-to-detector distances.

**Funding.** National Health and Medical Research Council, Australia (APP2011204), Australian Nuclear and Science Technology Organisation (AS233/IMBL/20286).

**Acknowledgments.** Part of this work was carried out on the Micro-Computed Tomography (MCT) beamline of the Australian Synchrotron, ANSTO. The authors are grateful to Microworks GmbH for producing the interferometer used in the present experiment, with a part of the work carried out with the support of the Karlsruhe Nano Micro Facility (KNMF, www.knmf.kit.edu), a Helmholtz Research Infrastructure at Karlsruhe Institute of Technology (KIT, www.kit.edu). The authors would like to thank Dr. A. Maksimenko of the Australian Synchrotron for assistance with the experiment, and Prof. D.M. Paganin of Monash University and Prof. L.J. Allen of the University of Melbourne for helpful discussions.

**Disclosures.** The authors declare no conflicts of interest.

**Data availability.** Data underlying the results presented in this paper may be obtained from the authors upon a reasonable request.